\documentstyle[preprint,aps]{revtex}

\begin{document}

\title{
Images and nonlocal vortex pinning in thin superfluid films} 
\author{H.H. Lee$^{1}$\cite{Lee}, J.M.F. Gunn$^{2}$\cite{Gunn} }
\address{$^{1}$ Department of Theoretical Physics, University of Manchester,}
\address{Manchester M13 9PL, United Kingdom.}
\address{$^{2}$ School of Physics and Space Research, University of Birmingham, 
Birmingham B15 2TT, United Kingdom.} 
 
\maketitle

\begin{abstract}
For thin films of superfluid adsorbed on a disordered substrate, we 
derive within a mean field 
(Hartree) description of the condensate the equation of motion for a 
vortex in the presence of a random potential. 
The compressible nature of the condensate leads to an 
effective pinning potential experienced by the vortex 
which is nonlocal, with a long range tail that smoothes out the random 
potential coupling the condensate to the substrate. 
We interpret this
nonlocality in terms of images, and relate the effective potential 
governing the dynamics
to the pinning energy arising from the expectation value of the Hamiltonian
with respect to the vortex wavefunction.
\end{abstract}

\section{Introduction}

Pinning of vortices in bulk helium has a long history (see for example Ref.\ 
\onlinecite{donn}) but the study of vortex pinning in thin films is at a much 
earlier stage of development: to our knowledge only one experimental study 
having been performed---Ref.\ \onlinecite{ellis}, although there are some 
aspects addressed in Ref.\ \onlinecite{adams}. In this paper we will 
concentrate on two dimensional pinning of idealised (in a sense defined presently) 
superfluid films which 
provides two simplifications compared to the situation in three dimensions: the 
ability to change the healing length as 
well as the obvious contrast between point vortices and extended vortex lines. 
The healing length is altered by changing the coverage of the Helium film, 
although quantifying this relationship is not easy.

Ellis and Li\cite{ellis} showed that by `swirling' a gold-plated mylar substrate a 
remanent vorticity could be created, where the density of pinned vortices was 
50,000cm$^{-2}$ so that the separation (45$\mu$) was of the order of $10^4$ 
larger than the film thickness (3.2nm). Putnam et al\cite{putn} have studied the topography of gold films deposited under similar conditions to those used by 
Ellis and Li; they find that the films have a surface of `rolling hills' with a 
characteristic length scale of 300\AA to 3000\AA. Thus the separation was 
considerably larger than the topographic features which are presumably the 
pinning agents. In this paper we will concentrate on the pinning of single 
vortices (and their dynamics) which should be appropriate under these 
experimental conditions. However we will confine ourselves to the simplest case 
of monolayer films.

We wish to derive from first principles the form of the pinning potential which 
a vortex experiences, given the potential in which the helium atoms move due to 
the substrate. We will see that these two quantities are not the same. We will 
describe the condensate at a mean field (Hartree) level using the hydrodynamic 
representation\cite{gross} (in terms of the two-dimensional density of the film 
and velocity potential of the flow) and of the resulting nonlinear Schr\"odinger 
equation (NLSE) which determines the condensate wave function\cite{noz}. We will
allow for some effects of compressibility---for instance that the 
(two-dimensional) density is modified to `screen' the random 
potential\cite{lee1}---and will find an equation of motion that bears some 
relation to the Magnus effect.

In this very thin film limit, there are two regimes in considering the behaviour of a vortex in a random potential. Firstly there is the region, near the centre 
of the vortex, where the largest contributions to the energy density are the 
kinetic energy of the fluid and the change in the films density that is caused 
by the flow. However at large distances, the dominant contribution to the energy 
density is the response of the density to the random (pinning) potential. Thus 
the problem falls into two parts and the manner in which to match the 
approximate solutions in those regions. The matching is performed by following 
the analysis of Neu\cite{Neu} which was constructed to determine the motion of 
a many vortex system in a compressible ideal fluid from first principles, 
reproducing the Kirchhoff result  
for widely separated vortices. There the two regions are: where 
single vortex effects are predominant (near the centre of each vortex) and 
where 
many vortex contributions are important (on longer length scales). Other 
expansions similar in spirit have been for compressible vortex 
rings\cite{roberts} and a number of works following Neu, which may be 
traced from Ref.\ \onlinecite{rub}.

In the monolayer regime, there is an additional complication: the 
superfluid-insulator transition (see, for example, the recent publications 
Ref.\ \onlinecite{crow} and Ref.\ \onlinecite{paz} and references therein). 
However sufficiently far away (a small fraction of a monolayer) from the 
onset of superfluidity, one may describe the helium film as being an `inert', 
nonsuperfluid, initial layer with a mobile, superfluid, film adsorbed on 
top\cite{lee}. The 
treatment in this paper will be of the latter part of the film, where the 
random potential is the residual one which includes any interaction with 
the inert layer.

The plan of the paper is as follows:
In section 2 we introduce the framework used to describe the vortex 
motion. Using a perturbation calculation, we derive the 
velocity of a vortex under the influence of an external potential, by
matching the \lq inner' and \lq outer' solutions in section 3.
In the next section, we establish that the dynamics are nonlocal in the
random substrate potential and explore some of the consequences. In the 
following section we relate the motion to the gradient of the expectation
value of the energy. In the final section we discuss several
issues which emerge from the rest of the paper and conclude.
 
\section{Notation and fluid representation of the NLSE}

We wish to study the influence of a disordered substrate qualitatively on a 
superfluid film. We therefore pick a very simple interaction between the bosons, 
namely a point interaction. 
Hence our starting point is the Hamiltonian
\begin{equation}
H = \sum_{i=1}^N -{{\hbar^2}\over{2m}}\nabla_i^2 + 
\frac{1}{2}\lambda \sum_{i\ne j}^N \delta({\mathbf r}_i - {\mathbf r}_j) 
+ \Delta \sum_{i=1}^N V({\bf r}_i)\label{Ham}
\end{equation}
Here the coordinates and Laplacian are two-dimensional, as we assume motion 
quantised normal to the substrate surface, with all atoms in the lowest 
state of that motion. $\Delta$ is the variance of the random two-dimensional 
potential $V$:
\begin{equation}
\Delta^2 = \langle [V({\mathbf r})]^2\rangle
\end{equation}
where the average is 
over the ensemble of potentials. Note that $V$ itself is dimensionless. 

We will treat the above Hamiltonian within the Hartree 
approximation, thus we will not include effects at the Bogoliubov approximation or 
beyond: for instance any analogue of the roton minimum. The Hartree approximation 
leads to the time-dependent NLSE which governs the condensate 
motion\cite{gross,noz}:
\begin{equation}
-\frac{1}{2}\nabla^2\phi+|\phi|^2\phi+\sigma V\phi=i{\partial\phi\over\partial 
t}\label{nlse}
\end{equation}
Where we are using appropriately scaled units: lengths are measured in units 
of the healing length, $\ell_{\rm h} = (\hbar^2/\lambda nm)^{1/2}$; 
$\sigma = \Delta/n\lambda$ is the dimensionless 
measure of the strength of the external potential, and energy is measured in units of 
$n\lambda$, which is the Hartree energy or chemical potential. $n$ is the average density of particles (equivalent to the 
condensate density in this Hartree case). $\phi$ is normalised to the size of the 
system, $\Omega$:
\begin{equation}
\int_{\Omega} |\phi|^2 \thinspace dV = \Omega
\end{equation}
so that in the absence of a potential, $\phi = 1$ everywhere within $\Omega$. 
Note that the speed of sound in the condensate, 
$(n\lambda/m)^{1/2}$,
is equal to unity with the above choice of units. 

To describe the dynamics of the condensate 
possessing a 
vortex structure, it is is natural to use the (Hartree)
fluid, or Madelung, representation\cite{gross}; by setting 
$\phi=\sqrt{\rho}\,e^{iS}$, Eqn.\ (\ref{nlse}) is equivalent to the pair of 
equations:
\begin{equation}
{\partial \rho\over \partial t}+\nabla \cdot(\rho\nabla S)=0
\label{cont}
\end{equation}
and
\begin{equation}
{\partial S\over \partial t}+\frac{1}{2}(\nabla S)^2+\rho+\sigma V
-\frac{1}{2}{\nabla^2\sqrt{\rho}\over \sqrt{\rho}}=0\label{bern}
\end{equation}
Eqns.\ (\ref{cont}) and (\ref{bern}) are respectively, 
the continuity, and Bernoulli equations describing 
the condensate flow. Density, $\rho$, is measured in units of $n$.

In the main body of this paper, we will be
interested in the behaviour of the flow field well beyond the core region, so 
that the rapid  variations in the 
condensate density and velocity field near the core itself are not considered.
Within this approximation we can neglect the \lq quantum pressure' 
term $\nabla^2\sqrt{\rho}/\sqrt{\rho}$ 
in the Bernoulli equation (\ref{bern}), (which we justify {\it a posteriori} presently) 
and use the set of equations: 
\begin{equation}
{\partial \rho\over \partial t}+\nabla\rho\cdot\nabla S+\rho\nabla^2S=0
\label{cont1}
\end{equation}
and 
\begin{equation}
{\partial S\over \partial t}+\frac{1}{2}(\nabla S)^2+\rho+\sigma V=0\label{bern1}
\end{equation}
For this work, an important solution to these equations is that of a vortex in two-dimensions:
\begin{equation}
S = \theta - t \quad \rho = 1 - \frac{1}{2} {{1}\over{r^2}}
\label{cleanvort}
\end{equation}
where $\theta$ is the polar angle. Note that the term $-t$ in the expression
for $S$ is the chemical potential, which cancels the term in the asymptotic density in the Bernoulli equation.

\section{Dynamics due to the substrate potential}

In the Magnus equation for incompressible
point vortices, the presence of a potential coupling to the vortex will change 
its motion from being determined completely by the value of the velocity field 
at the vortex centre. (The latter result, due to Kelvin and Helmholtz, has been 
derived in the context of Landau-Ginzburg theory by Neu\cite{Neu}.) In this section 
we will derive an approximate solution for a vortex in the presence of 
a potential which couples to the helium atoms in the film. This will allow us to 
derive in the next section the form of the potential which applies to the 
{\it vortex}, as against the helium atoms, and hence appears in the equation of 
motion of the vortex. We apply Neu's\cite{Neu} method of matched asymptotic 
expansions in this section.

We will perform a linear response calculation using the fluid (Madelung) representation of the condensate wave function. This is preferable to using linear response theory directly - in terms of the condensate wave function and the Bogoliubov excitations in the presence of the vortex. In the latter approach we would need to calculate the matrix elements of, for instance, the random potential between the ground state and the excitations. Since the wave functions of the excitations around a vortex are only known asymptotically this would not be straightforward.

Before embarking on any calculation, we wish to divide the problem into one 
of `inner' and `outer' solutions. In the inner region, the vortex motion is 
dominant, in terms of both density and velocity field, and in the outer region, 
the random potential is dominant.
We can estimate the boundary between these two regions to be where the 
perturbation in the density due to flow around the vortex is equal to the 
perturbation to the 
density due to the random potential; namely at to be a distance $r_c$ where:
\begin{equation}
{1\over r_c^2}\sim \sigma\,\,\Rightarrow\,\,r_c\sim\sigma^{-1/2}
\label{bound}
\end{equation}
and $r_c$ is measured in units of the healing length.

The inner solution can be determined 
by perturbation theory on Eqns.\ (\ref{cont1}) and (\ref{bern1}) in powers 
of $\sigma$, since  
the random potential is assumed to be weak.

For the outer solution, we want to rescale the position and time 
variables ${\mathbf r}$ and $t$ to make the dominance of $\sigma$ at large 
distances manifest. We make the substitutions:
\begin{equation}
{\mathbf r}\to{\mathbf r}'=\sigma{\mathbf r}\,,\ \,\,t\to t'=\sigma^2 t
\label{scale}
\end{equation}
In terms of these new variables, the continuity and Bernoulli equations of 
(\ref{cont1}) and (\ref{bern1}) become (dropping primes):
\begin{equation}
{\partial \rho\over \partial t}+\nabla\rho\cdot\nabla S+\rho\nabla^2S=0
\label{conts}
\end{equation}
and
\begin{equation}
\sigma^2{\partial S\over \partial t}+\sigma^2
\frac{1}{2}(\nabla S)^2+\rho+\sigma V=0
\label{berns}
\end{equation}
The continuity equation remains unchanged but the Bernoulli equation is 
modified---we may neglect the time-dependence of $S$ (apart from the 
term $-t/\sigma^2$ which provides the chemical potential, as 
mentioned below Eqn.\ (\ref{cleanvort}).
In the outer region we expand the density and velocity potential as:
\begin{equation}
\rho=1+\sigma\rho_1({\mathbf r})\,,\ \,S=S_0({\mathbf r}-{\mathbf R})
+\sigma S_1({\mathbf r})
\label{expn}
\end{equation}
Note that the density does not depend on the position of the vortex, 
as the terms in Eqn.\ 
(\ref{berns}) which depend on $\mathbf R$ are second order in $\sigma$, 
similarly we may neglect the time-dependence of $\rho$ as time does 
not enter  Eqn.\ 
(\ref{berns}) to first order in $\sigma$. In Eqn.\ (\ref{expn}) the 
zeroth order flow field 
retains a vortex flow, 
so that $\oint\nabla S_0\cdot d{\mathbf r}=2\pi$,
despite a  
distortion of the flow as a whole due to the linear addition of 
$\nabla S_1$.

Substituting the expansion (\ref{expn}) into Eqn.\ (\ref{conts}) and 
(\ref{berns}) we have 
to $O(\sigma)$:
\begin{eqnarray}
\rho_1&=&-V \nonumber \\
\nabla^2 S_1+\nabla S_0\cdot \nabla\rho_1&=& 0 \label{firsto}
\end{eqnarray}
Solving for $S_1$ and denoting the outer solution by $S^{>}_1$, we find:
\begin{equation}
S_1^{>}({\mathbf r})=\int G({\mathbf r}-{\mathbf r}')\,\bbox{\gamma}\cdot
\nabla V({\mathbf r}')\,
d{\mathbf r}'
\label{oneout}
\end{equation}
where
\begin{equation}
G({\mathbf r}-{\mathbf r}')={1\over 2\pi}\log|{\mathbf r}-{\mathbf r}'|
\label{green}
\end{equation}
and we have introduced the more compact notation 
$\bbox{\gamma}\equiv \nabla S_0({\mathbf r}-{\mathbf R})$ for the vortex field.
It is important to note that the variables here are the {\it scaled} 
ones defined in Eqn.\ (\ref{scale}).

Turning to the inner solution, we revert to the unscaled variables, and
solve the original equations (\ref{cont1}) and (\ref{bern1}). 
We require the solution to be matched to 
the outer one at distances much greater than the core radius, i.e. at
dimensionless distances $r\gg 1$.
In the inner region we can write to $O(\sigma)$:
$$\rho=\rho_0({\mathbf r}-{\mathbf R})+\sigma\rho_1\,,\,\,\,\,\,\,
S=S_0({\mathbf r}-{\mathbf R})+\sigma\nabla S_1
$$
with $\rho_0$ given by
\begin{eqnarray}
\rho_0&=&1-{1\over 2|{\mathbf r}-{\mathbf R}|^2}\nonumber\\
&=&1 - \delta\rho_0({\mathbf r})\nonumber\\
\label{rho0}
\end{eqnarray}
So $\delta\rho_0$ is the same as the change in density due to the vortex in the 
absence of the random potential. The fluid equations taken to $O(\sigma)$ are 
then:  
\begin{eqnarray}
\rho_1&=&-V+\bbox{\gamma}\cdot ({\dot {\mathbf R}}-\nabla S_1) \nonumber \\
\rho_0\nabla^2S_1&+&\bbox{\gamma}\cdot\nabla\rho_1+\nabla\rho_0\cdot(\nabla
S_1-{\dot {\mathbf R}})=0\label{innerone}
\end{eqnarray}
The eventual aim is to find ${\dot {\mathbf R}}$ in terms of  
$S_1$; that is,  we look for the change in flow caused by the 
potential, which then determines the vortex velocity through advection.
Examining Eqn.\ (\ref{innerone}), we can inspect the individual terms to see 
which ones 
will dominate in the ${\mathbf r}\to\infty$ limit for the matching.
Eliminating $\rho_1$, and omitting terms of order $1/r^3$ and higher,
we are left with
$$\nabla^2 S_1=\bbox{\gamma}\cdot\nabla V$$
which can be solved $S_1^{<}({\mathbf r})$ for to obtain
$$S_1^{<}({\mathbf r})=\int G({\mathbf r}-{\mathbf r}')\,\bbox{\gamma}\cdot\nabla 
V({\mathbf r}')\,d{\mathbf  r}'\,+O(1/r)$$
where $S_1=S^{<}_1$ denotes the inner solution.

We will now match the inner and outer solutions at a distance 
$r\sim \sigma ^{-\alpha}$, with $\alpha>0$ chosen to ensure corrections are 
small. Firstly, we must consider the {\it outer} solution back in terms of 
the {\it unscaled} variables. But from Eqn.\ (\ref{oneout}), it can be seen 
that there is 
in fact no change in its form, as lengths only enter $S^{>}_1$ and the scaling 
of $\bbox{\gamma}\cdot\nabla$ and the element of integration $d{\mathbf r}$ cancel.
Hence we can conclude that matching is {\it perfect} up to $O(\sigma)$. 
At $r\sim\sigma^{-\alpha}$, neglected terms in the equation 
determining the inner $S_1$ are $1/r$ or a factor of 
$\sigma^{\alpha}$ smaller; thus any inconsistency is negligible. We will 
make a more refined consistency check presently.

Note that (neglecting corrections) $S_1$ can be integrated by parts to give:
\begin{equation}
S_1({\mathbf r})=-\int \bbox{\gamma}({\mathbf r}')\cdot \nabla G({\mathbf r}-
{\mathbf r}')\,
V({\mathbf r}')\,d{\mathbf r}'
\label{innerone1}
\end{equation}
where we assume that the \lq surface' integral vanishes (and $\nabla \cdot 
\bbox{\gamma} =0$).
This alternate form for $S_1$ will be useful in the next section.

So far the velocity of the vortex, ${\dot {\mathbf R}}$, has not been determined. 
To do so, 
we Taylor expand the outer solution (\ref{oneout}) 
about ${\mathbf r}'={\mathbf R}'$ and check that 
corrections to this expansion are negligible in the matching region. This 
ensures that the Taylor expansion is consistent with the matching\cite{Neu}.
Thus we have
$$S^{>}({\mathbf r}')=S^{>}({\mathbf R}')+({\mathbf r}'-{\mathbf R}')\cdot\nabla 
S^{>}_1({\mathbf R}')+O({\mathbf r}'^2)$$
In the second term on the right, $\nabla S^{>}_1$ appears as 
$\nabla S^{>}_0$ is perpendicular to $({\mathbf r}'-{\mathbf R}')$.

We now appeal to the Helmholtz 
theorem which states that the vorticity moves with the local velocity field (the 
\lq dynamical boundary condition'). 
Using Eqn.\ (\ref{oneout}) this gives the result for ${\dot {\mathbf R}}$ as:
\begin{equation}
{\dot {\mathbf R}}=\int \nabla_{\mathbf R} G({\mathbf R}-{\mathbf r})\,
\bbox{\gamma}\cdot\nabla V({\mathbf r})\,d{\mathbf r}
\label{vorvel}
\end{equation}
where $\nabla_{\mathbf R}\equiv \partial /\partial {\mathbf R}$.

To check consistency, we examine the size of 
$({\mathbf r}'-{\mathbf R}')\cdot{\dot {\mathbf R}}=\sigma({\mathbf r}-
{\mathbf R})\cdot{\dot {\mathbf R}}$ compared 
with the corrections $1/r$ and $\sigma^2r^2$ (i.e. the $O(r'^2)$ in the Taylor 
expansion).
We require $\sigma r\gg 1/r$ and $\sigma^2r^2$ for $r\sim\sigma^{-\alpha}$.
The first condition gives $\sigma^{(1-\alpha)}\gg\sigma^{\alpha}\Rightarrow
\alpha>1/2$, while the second one gives $\sigma^{(1-\alpha)}\gg
\sigma^{(2-2\alpha)}\Rightarrow \alpha<1$.
Hence for both sets of corrections to be negligible, we need:
$$1/\sigma^{1/2}<r<1/\sigma$$
Since this region does exist, the expansion is consistent.

\section{Interpretation of vortex motion in terms of images}

We have now deduced the form for ${\dot {\mathbf R}}$ which governs 
the vortex motion in an external potential, and in this section consider 
some of the consequences and the interpretation of Eqn.\ (\ref{vorvel}).
Firstly, note that ${\dot {\mathbf R}}$ is clearly related to the potential in 
a {\it nonlocal} manner.
Using Eqn.\ (\ref{innerone1}), the expression for ${\dot {\mathbf R}}$ in 
Eqn.\ (\ref{vorvel}) can be rewritten as:
\begin{equation}
{\dot {\mathbf R}}=-\nabla_{\mathbf R}\int\bbox{\gamma}({\mathbf r})\cdot
\nabla G({\mathbf R}-{\mathbf r})
V({\mathbf r})\, d{\mathbf r}
\label{rdote}
\end{equation}
We can further manipulate this by using the identity
$\bbox{\theta}\cdot\nabla V=({\hat {\mathbf z}}\times \nabla V)\cdot
\hat {\mathbf r}$ where ${\hat {\mathbf z}}$ is perpendicular to the plane of motion,
so that (\ref{rdote}) becomes:
$${\dot {\mathbf R}}=-{\hat {\mathbf z}}\times
\int(\bbox{\mu}\cdot\nabla_{\mathbf R})\nabla_{\mathbf R}G({\mathbf R}
-{\mathbf r})V({\mathbf r})\,d{\mathbf r}$$
with $\bbox{\mu}\equiv ({\mathbf R}-{\mathbf r})/|{\mathbf R}-{\mathbf r}|^2$.

This implies:
$${\hat {\mathbf z}}\times{\dot {\mathbf R}}=\nabla_{\mathbf R}V_{{\text {eff}}}$$
where 
\begin{equation}
V_{{\text {eff}}}({\mathbf R})={1\over 2\pi}
\int{V({\mathbf r})\over |{\mathbf R}-{\mathbf r}|^2}\,d{\mathbf r}
\label{vefft}
\end{equation}
This reproduces the form of the Magnus equation but with an 
effective potential $V_{{\text {eff}}}$ which is nonlocal in the substrate 
potential with a $1/|{\mathbf R}-{\mathbf r}|^2$ kernel.

The most natural interpretation of this result is in terms of images. To 
indicate the relevance of images, note that if the potential were sufficiently 
large then the density of the film would become zero. For simplicity assume that 
the boundary of the film is a straight line. In that case in a standard manner 
we may take into account the boundary condition that the velocity field 
perpendicular to the film boundary is zero on the boundary by adding the 
velocity field of an image vortex. It is plausible that when the potential 
causes changes in the density, but is not sufficient to drive the density to 
zero, there will still be features in the velocity field which may be 
attributed to images. Their presence ensures that the flow field obeys the 
continuity equation with terms involving $\nabla \rho$. We will now indicate, 
in terms of a simple example, how Eqn.\ (\ref{vefft}) is interpretable in 
such a manner.

We now give an example of vortex motion for the case 
of a repulsive delta function potential centred at ${\mathbf r}_a$:
\begin{equation}
V({\mathbf r})=\delta({\mathbf r}-{\mathbf r}_a)
\label{delta}
\end{equation}
The Greens function (\ref{green}) satisfies the identity:  
\begin{equation}
\nabla\nabla G({\mathbf r})=-{1\over \pi r^2}
\lbrace \cos 2\theta ({\mathbf {\hat e}}_x{\mathbf 
{\hat e}}_x-{\mathbf {\hat e}}_y{\mathbf {\hat e}}_y)+
\sin 2\theta({\mathbf {\hat e}}_x{\mathbf {\hat e}}_y+
{\mathbf {\hat e}}_y{\mathbf {\hat e}}_x)\rbrace
\label{nabnab}
\end{equation}
Using Eqns.\ (\ref{delta}) and (\ref{nabnab}), the vortex velocity 
(from (\ref{rdote})) is given by:
\begin{equation}
{\dot {\mathbf R}}={\bbox{\theta}_a\over |{\mathbf R}-{\mathbf r}_a|^3}
\label{dipole}
\end{equation}
where $\bbox{\theta}_a$ is perpendicular to ${\mathbf r}_a-{\mathbf R}$.

To interpret this result, let us compare an image approach to the motion of
a vortex in the presence of an impenetrable circular region centred at 
${\mathbf r}_a$. 
Then the magnitude of the velocity experienced by the vortex due to the image 
is of $O(1/d^3)$ where $d$ is the 
distance between the vortex and its image (this is because the magnitude of the 
velocity field due to the vortex itself at the circle is $o(1/d)$ and hence the 
dipole moment required to cancel out the normal component is of this magnitude, 
hence the velocity field back at the vortex will be $O(1/d \times 1/d^2$). This 
has the same functional form as Eqn.\ (\ref{dipole}), leading to the 
interpretation in terms of a dipole image at the delta function potential, but 
with a strength which is not unity but proportional to the strength of the 
potential. This has some analogy with the difference between images in 
electrostatics in the case of metals and dielectrics. The direction of the 
motion of the vortex is also consistent: in the image picture the vortex would 
be advected around the circle, in agreement with the angular unit vector in 
Eqn.\ (\ref{dipole}). It is easy to check that a line of delta functions 
produces motion parallel to the line (of magnitude $1/d$), in analogy with the 
motion of a vortex in the presence of a wall.

We note parenthetically that the response of the condensate to a delta function perturbation without a vortex is of a much simpler form: a delta function change of the condensate. This is because the repsonse of the condensate to a perturbation relaxes on a scale of the healing length; the fluid description only represents behaviour on scales larger than the healing length, hence the perturbation is of a delta function form within the fluid approximtation. However as we have seen, there is an interesting and nontrivial response on length scales in excess of the healing length in the presence of a vortx.

Having established that the vortex obeys a Magnus equation of motion and hence 
moves parallel to the equipotentials of the effective potential, 
$V_{{\text {eff}}}$, we may 
immediately draw some additional conclusions using the work of Trugman and 
Doniach\cite{Trug}. 
A consequence of 
Eqn.\ (\ref{vefft}) is that the 
vortex trajectories must be closed, with $V_{{\text {eff}}}$ instead of $V$ as 
the underlying (local) potential. In other words, the vortex travels along the
equipotential lines of the former. 
The exception to this is at the saddle points
of the effective potential where  the vortex may \lq percolate', allowing in
principle an extended orbit across the system. (One can think of 
of \lq hills' and \lq valleys' representing the potential,
and the percolation threshold corresponding to \lq lakes' which fill
up the valleys, to connect). However, the
speed of the vortex goes to zero (due to the vanishing of the gradient of the
potential) at the saddle points, so that the vortex transport is in fact
pathological. These points have immediate parallels with the guiding centre 
motion of charged particles in the quantum Hall effect\cite{tsu,trug1,jans}, 
where the existence of percolating paths is related to the mobility edge 
believed to occur in the middle of a Landau sub-band. Obviously the application 
of these ideas to vortices would only be appropriate when the vortex density 
was extremely low.   

\section{Pinning energy}

So far the discussion has been in terms of the equations of motion. In this 
section we show that we may calculate the pinning energy and moreover the force
which occurs in the equation of motion of the vortex is the gradient of that 
potential.

The expectation value of the energy is:
$$E = \frac{1}{2} \rho (\nabla S)^2 + \frac{1}{2} \rho^2 + \sigma V \rho$$
Now, substituting in the expansions for the density and velocity potential 
to $O(\sigma)$,
there are four terms which depend on the vortex position contributing to this energy 
(here $\delta \rho_0({\mathbf r})$ is defined in Eqn.\ (\ref{rho0})): 
\begin{equation}
\Sigma=\int d{\mathbf r}\,
[\delta \rho_0V+\rho_1\delta\rho_0+\rho_1\frac{1}{2}\gamma^2
+\bbox{\gamma}\cdot\nabla S_1]
\label{pineng}
\end{equation}
where we have taken the vortex core to be situated at the origin.

We can interpret the separate terms on the right hand side of this equation 
as follows.
The first term is just the
pinning  potential energy associated with the decreased 
density as the core is approached. The second term is the interaction of this 
decrease in the density with the distortion of the condensate due to the 
substrate potential. The third term comes from the 
change in the kinetic energy due to the substrates effect on the density. The 
final term comes from the change in the kinetic energy due to the distortion of 
the flow field due to the substrate (i.e. 
the consequence of the vortex flowing around \lq obstacles' in the 
substrate). Now the first and second terms cancel as $\rho_1 = - V$. Thus we 
are left with the final two terms:
\begin{equation}
\Sigma=\int d{\mathbf r}\, [-V\frac{1}{2}\gamma^2
+\bbox{\gamma}\cdot\nabla S_1]
\label{pineng1}
\end{equation}
The (inverse) power law nature of $\delta \rho_0$ means that the pinning is 
{\it nonlocal} as to be expected from the discussion in section 4. We may 
now rewrite the second term using the two-dimensional divergence theorem:
\begin{eqnarray}
\int d{\mathbf r} \,\bbox{\gamma} \cdot \nabla S_1 &=& \int \nabla \cdot 
(\bbox{\gamma}S_1)  - \int d{\mathbf r}\,\nabla\cdot \bbox{\gamma} S_1 
 \nonumber \\
&=& \int d\bbox{\mathcal S} \cdot \bbox{\gamma} S_1 - \int d{\mathbf r}\,\nabla\cdot 
\bbox{\gamma} S_1  \nonumber \\
&=& \int d\bbox{\mathcal S} \cdot \bbox{\gamma} S_1 - 0 \label{divth}
\end{eqnarray}
as $\nabla \cdot \bbox{\gamma} =0$. 

The surface integral in Eqn.\ (\ref{divth}) will be zero as the velocity field 
parallel to the boundary will vanish due to additional images (which we do not 
discuss explicitly). The final result is thus:
$$E = - \int \frac{1}{2} \gamma^2 V  \,d{\mathbf r}$$
i.e. the interaction of the depletion due to the centrifugal forces interacting 
with the substrate potential.

It should be stressed that the core region, $r<1$, will contribute to the 
pinning energy a term of comparable magnitude to the long range part which we 
are explicitly discussing; however the latter yields only a short range 
correlated potential as against the long range correlations which the above has. 
It is readily checked that minus the gradient of this potential is indeed the 
term on the r.h.s of the equation of motion of the vortex and hence is 
consistent with our considerations of the equation of motion.

\section{Discussion and conclusions}

In this paper we have only discussed the motion of a single vortex in the
presence of a random potential. Let us now briefly discuss the relevance of
this to the more realistic case where there are many vortices present in the
system. The vortices may be the remanent vorticity which is pinned by the
random potential, or it may be that the substrate is rotating and hence there
is some vorticity with a net sign.

We concentrate on the case where the disorder is strong compared to the
interaction of the vortices at the average separation. Of course in the case of
a rotating substrate the
logarithmic nature of the interaction between the vortices will ensure that
the density does not fluctuate too much, however the local order in the
vortices is determined by the higher Fourier components of the random
potential, and the shear modulus of the vortex lattice which will decrease with
density.

Because of the long-range interactions we must consider the possibility of an
analogue of a Coulomb gap which occurs for electronic excitations in a highly
disordered system\cite{efros,cgbook,fisher}. The `gap' refers to the density
of states for adding an extra vortex to the system. In the case of vortices
with a logarithmic interaction, the density of states, $n(\epsilon)$, as 
the energy,
$\epsilon$, tends to zero, behaves as $n(\epsilon)\sim \epsilon^3$. However the 
situation with
excitations where a vortex is moved within the system is more complicated as one 
is creating a dipole and the density of these excitations does not tend to zero 
at zero energy. The
above results all assume that the particles (vortices in our case) reside on a
lattice; however if the potential is smooth then as well as excitations where
the vortex is moved between local minima in the overall potential (due to the
combination of the underlying potential and the effect of the other vortices)
there are excitations where it remains in the minimum which it started in. It is 
those excitations with which we have been concerned in this paper.

These excitations would (at least in principle) 
contribute to the damping of a torsion balance. This
may be seen readily by considering a vortex in an approximately harmonic
potential well ($V_{\rm eff}({\bf r}) \simeq (1/2) K r^2$) 
experiencing a superflow with amplitude $A$ and frequency 
$\Omega$ in the direction ${\bf {\hat x}}$. The equation of motion for the
vortex with position $\bf r$ and unit circulation is then:
$$({\bf {\dot r}} - A \cos\Omega t {\bf {\hat x}}) \times {\bf {\hat
z}} = - K {\bf r}$$
The effect of the superflow is to give a response in terms of the motion of the
vortex which is resonant if $\Omega = \omega_0 = K$. (That is if the
period of the superflow is the same as the frequency of the vortex moving
around the well.) For instance the
$x$-component of the vortex position is:
$$x = \alpha\cos \omega_0t + \beta \sin  \omega_0 t +
{{A\Omega}\over{\Omega^2-\omega_0^2}}\sin \Omega t$$
So those vortices  which have associated natural
frequencies which are the same as that of the torsion balance will be excited
significantly. Of course this frequency is rather low on a microscopic scale
and so it may be insignificant in practice.

The extent to which the excitations are single-vortex in their nature is hard to
estimate in general. However one may start from the assumption of only a single
vortex is excited and then examine whether the amplitude spreads to other
vortices. The single vortex excitation has a characteristic frequency involved
in precessing around an equipotential of $V_{\mathrm eff}$. Whether this may
then excite neighbouring vortices depends on whether the natural frequencies of
the neighbouring vortices are sufficiently similar for a `resonant' process. If
the random potential has a sufficiently large variance, as is assumed here,
this is very unlikely so most excitations will only involve one vortex. Hence 
the results of this paper will be relevant.

We have not considered here the effect of the vortices having a nonzero
inertial mass. The effect of this may be deduced by analogy with the case of a
charged particle in a magnetic field---which is very similar upon the addition
of an inertial mass term. As well as the `guiding centre' motion of the vortex
along the equipotentials, there is now some (fast) `cyclotron motion'. (A
discussion of this effect in classical hydrodynamics is given by
Lamb\cite{lamb}). However the value of the inertial mass (and hence the
magnitude of the cyclotron motion) has been a controversial issue. The
suggestion of Baym and Chandler\cite{baym} was that it was associated with the
virtual mass due to the backflow around a cylinder of radius approximately
the coherence length. However more recently Duan\cite{duan} has argued that the
mass is larger by a factor of 20-30 due to the finite compressibility of
helium. In terms of the analogue of the `cyclotron radius', $\ell_{\rm m}= 
(\rho/\mu)^{-1/2}$, where $\mu$ is the mass of the vortex, will only change
from being roughly an interatomic distance to being 4-5 times that size.

In this paper we have nothing to say about corrections to the mean field
picture. This has been addressed in part by Niu et al.\cite{niu}, where the
magnitude of the vortex effective mass has been considered from a quantum 
point of view. The relationship between these calculations and the effects due to 
backflow is not very clear and we will make no more comment on these matters.

In summary we have derived the effective potential that a vortex experiences 
due to a potential coupled to the underlying bosons. The former is much smoother
than the latter, with long range tails. We showed that the tails may be interpreted as 
due to `images' caused by variations in the condensate density.  Moreover
the effective potential was equal to the pinning energy in the 
expectation value of the Hamiltonian with a Hartree wave-function representing the vortex.

\section*{Acknowledgments}

We would like to acknowledge the support of the EU through Grant No. CT90 0020. 
HHL would like to thank Prof. M.A. Moore for helpful discussions, and
the SERC for support while this work was carried out.

\end{document}